\documentclass{iau}
\usepackage{graphicx}

\title[The COBRA Survey] %% [give here short title] %%
{Extragalactic Jets as Probes of Distant Clusters of Galaxies and the \\ Clusters Occupied by Bent Radio AGN (COBRA) Survey}

\author[E. L. Blanton et al.]   %% [give here the short author list; use "et al." if 3 authors or more] %%
{
Elizabeth L. Blanton$^1$,
Rachel Paterno-Mahler$^1$,
Joshua D. Wing$^2$,
M. L. N. Ashby$^2$,
Emmet Golden-Marx$^1$,
Mark Brodwin$^3$,
E. M. Douglass$^4$,
Scott W. Randall$^2$,
T. E. Clarke$^5$
}

\affiliation{
$^1$Institute for Astrophysical Research and Department of Astronomy, Boston University \\ 
725 Commonwealth Avenue, Boston, MA 02215 USA \\ 
email: {\tt eblanton@bu.edu, rachelpm@bu.edu, emmetgm@bu.edu} \\
[\affilskip]
$^2$Harvard-Smithsonian Center for Astrophysics \\ 
60 Garden Street, Cambridge, MA 02138 USA \\
email: {\tt jwing@cfa.harvard.edu,mashby@cfa.harvard.edu,srandall@cfa.harvard.edu} \\
[\affilskip]
$^3$Department of Physics and Astronomy, University of Missouri \\ 
Kansas City, MO 64110 USA \\
email: {\tt brodwinm@umkc.edu} \\
[\affilskip]
$^4$Department of Physics, Farmingdale State College \\ 
Lupton 200, 2350 Broadhollow Rd., Farmingdale, NY 11735 USA \\
email: {\tt edmund.douglass@farmingdale.edu} \\
[\affilskip]
$^5$Naval Research Laboratory \\ 
Code 7213, 4555 Overlook Ave SW, Washington, DC 20375 USA \\
email: {\tt tracy.clarke.ca@nrl.navy.mil}
}

\pubyear{2014}
\volume{313} 
\pagerange{xx--xx}
\jname{Extragalactic jets from every angle}
\editors{F. Massaro, C.C. Cheung, E. Lopez, A. Siemiginowska, eds.}
\begin{document}

\maketitle

\begin{abstract}
We are conducting a large survey of distant clusters of galaxies using radio sources with bent jets and lobes as tracers. 
These radio sources are driven by AGN and achieve their bent morphologies through interaction with the surrounding gas found in 
clusters of galaxies. Based on low-redshift studies, these types of sources can be used to identify 
clusters very efficiently. We present initial results from our survey of 653 bent-double 
radio sources with optical hosts too faint to appear in the SDSS. The sample was observed in the infrared with {\it Spitzer}, 
and it has revealed $\sim$200 distant clusters or proto-clusters in the redshift range $z\sim0.7 - 3.0$. The sample of 
bent-doubles contains both quasars and radio galaxies enabling us to study both radiative and kinetic mode feedback in cluster 
and group environments at a wide range of redshifts.
\end{abstract}

\firstsection % if your document starts with a section,
\section{Introduction}

Distant clusters are important cosmological probes as well as laboratories for studying the formation and evolution of galaxies.
High-redshift clusters are very difficult to find -- the faint optical magnitudes of the associated galaxies and the X-ray
emission from the intracluster medium (ICM) will often be missed by current surveys at redshifts of $z > 1 - 1.5$. Any clusters that
are revealed at these types of distances add significantly to the few that are known. Using a unique method employing radio sources
as tracers, we have confirmed the existence of a large number of clusters with redshifts as high as $z=0.96$ 
(\cite[Blanton et al. (2000)]{Blanton_etal00},
\cite[Blanton et al. (2001)]{Blanton_etal01},
\cite[Blanton et al. (2003)]{Blanton_etal03},
\cite[Wing \& Blanton (2011)]{Wing_11}).

Clusters of galaxies usually contain bright elliptical or cD galaxies near their centers.  These galaxies often host
a supermassive black hole in their cores.  Accreting supermassive black holes are the origin of active-galactic nuclei
(AGN) and can produce jets and lobes detected in the radio that have a profound effect on the surrounding X-ray-emitting intracluster medium (ICM).
This AGN feedback transports energy and magnetic fields into
clusters.  AGN contribute to regulating star formation, and limit the total masses of galaxies.  They affect scaling relations in clusters,
such as the relationship between X-ray luminosity and temperature.  The ICM affects the radio lobes by confining and distorting them.

Certain types of radio sources are more often found in clusters than others,
and these sources can be used as tools to discover clusters at a wide range of redshifts which can be used for studies of  
galaxy formation and evolution, as well as cosmology.  
Bent, double-lobed (including wide-angle-tail [WAT]) radio sources are particularly
good tracers for clusters of galaxies at a range of redshifts.
The lobes are likely bent as the result of the relative motion between the radio galaxy and the dense ICM which provides ram pressure to 
bend the lobes.  Since these sources are often associated with central cD or giant elliptical galaxies with small peculiar velocities
relative to the cluster mean, they may sometimes be found in merging clusters where the ICM has been set in motion, or in relatively
relaxed systems exhibiting ``sloshing'' of the central ICM 
(\cite[Ascasibar \& Markevitch (2006)]{Ascasibar06},
\cite[Mendygral et al. (2012)]{Mendygral_etal12})
related to the cluster minor merger history.
Bent-double radio sources can be found in clusters and groups with a wide range of masses. Our survey, therefore, is an important complement
to other cluster surveys that select only more massive clusters at high-z.

Using data at optical, near-infrared (NIR), mid-infrared, X-ray, and radio wavelengths, we are examining the cluster 
environments of radio sources at high redshifts.  We have laid significant groundwork by studying these sources in detail at
low-to-moderate redshifts 
(\cite[Blanton et al. (2000)]{Blanton_etal00},
\cite[Blanton et al. (2001)]{Blanton_etal01},
\cite[Blanton et al. (2003)]{Blanton_etal03},
\cite[Wing \& Blanton (2011)]{Wing_11},
\cite[Wing \& Blanton (2013)]{Wing_13},
\cite[Douglass et al. (2008)]{Douglass_etal08},
\cite[Douglass et al. (2011)]{Douglass_etal11},
\cite[Douglass (2012)]{Douglass12},
\cite[Paterno-Mahler et al. (2013)]{PaternoMahler_etal13},
\cite[Douglass et al. (2014)]{Douglass14}).
Our sample of high-redshift clusters will be used to help determine the epoch of galaxy formation, study the galaxy populations
of clusters over time, and place constraints on cosmological models including dark energy. By the nature of their selection, all of the clusters
will contain radio-loud active galaxies or quasars. We will therefore study AGN feedback at high-z, including ``radio'' or ``kinetic'' mode feedback
and ``quasar'' or ``radiative'' mode feedback 
(\cite[Fabian (2012)]{Fabian12}).
Poor clusters at high-z are particularly important for 
studying AGN feedback 
(\cite[Mittal et al. (2011)]{Mittal_etal11},
\cite[Puchwein, Sijacki, \& Springel (2008)]{Puchwein08}).

\section{Groundwork}

We have laid significant groundwork for our high-redshift cluster survey by studying the environments of radio sources at low-to-moderate
redshifts. We have analyzed the cluster environments of 12 low-z clusters containing bent, WAT radio sources using archival {\it Chandra}
data 
(\cite[Douglass (2012)]{Douglass12},
\cite[Douglass et al. (2014)]{Douglass14}).
In addition, two comparison samples of similar size were analyzed using the same analysis methods
-- a sample of known cool core clusters and a sample of known non-cool core, merging clusters. Properties of the comparison sample clusters were clearly
dellineated. We defined six cool core properties (e.g. central temperature drop, peaked central density profile) and found that all clusters
in the cool core comparison sample displayed at least five of these properties while all clusters in the non-cool core, merger sample
displayed one or fewer of these properties. The clusters in the WAT sample were much more heterogeneous -- four had at least five cool core
properties, three had two or three cool core properties, and five showed one or fewer cool core properties. Therefore, bent, WAT sources
may appear in a wide range of environments from related cool cores to large scale cluster-cluster mergers. Examples of two extremes of these
environments are shown in Fig.\ 1. A562 is shown in the left panel 
(\cite[Douglass et al. (2011)]{Douglass_etal11})
and is undergoing a large scale merger. A2029,
in the right panel 
(\cite[Paterno-Mahler et al. (2013)]{PaternoMahler_etal13})
is a well-known cool core cluster that is relaxed on large scales. However, there is evidence
for sloshing of the cluster core gas related to a past off-axis merger. This sloshing gas (the signatures of which can persist for Gyr)
is likely bending the radio lobe of the source in the center of A2029.

In addition to examining the X-ray cluster environments of a small sample of bent, double-lobed radio sources, we have studied the 
optical environments of much larger radio source samples using data from the Sloan Digital Sky Survey 
(SDSS, \cite[Abazajian et al. (2009)]{Abazajian09}) 
(\cite[Wing \& Blanton (2011)]{Wing_11},
\cite[Wing \& Blanton (2013)]{Wing_13}).
Samples of bent-double lobed radio sources were selected from the VLA Faint Images of the Radio Sky at Twenty-cm (FIRST) survey
(\cite[Becker, White, \& Helfand (1995)]{Becker95}), including
all bent three-component sources (including a core and two lobes) contained in the survey. We included sources that were identified visually
as well as from a pattern recognition program 
(\cite[Proctor (2006)]{Proctor_06}) --
the visual bent and auto bent samples, respectively. For comparison, we also
examined the environments of straight three-component radio sources and single-component radio sources. Owing to the magnitude limits of
SDSS, we were able to identify luminous elliptical galaxy hosts to $z\approx0.7$ and study the cluster environments (dominated by fainter
galaxies) to $z\approx0.5$. Through this work, we found hundreds of clusters associated with the bent radio sources out to $z\approx0.5$.
In addition, we found that the bent radio sources were most often found in clusters (with association rates up to 80$\%$ depending on the
sub-sample), followed by the extended, straight sources and finally, the single-component sources 
(\cite[Wing \& Blanton (2011)]{Wing_11}).
The distribution of redshifts for the optical hosts for the radio source samples is shown as $r-i$ color vs.\ redshift in the left panel of 
Fig.\ 2. In addition to the elliptical galaxy hosts, a small number of sources were associated with high$-z$ quasars allowing their redshifts to be measured up to $z\approx3$ (beyond the limit displayed here) in the SDSS. An example cluster associated with a radio source at $z=0.25$ is shown in the right panel of Fig.\ 2
(\cite[Wing \& Blanton (2011)]{Wing_11}).

\begin{figure}[t]
\vskip 2.8truein
\begin{center}
%\hbox{
\includegraphics{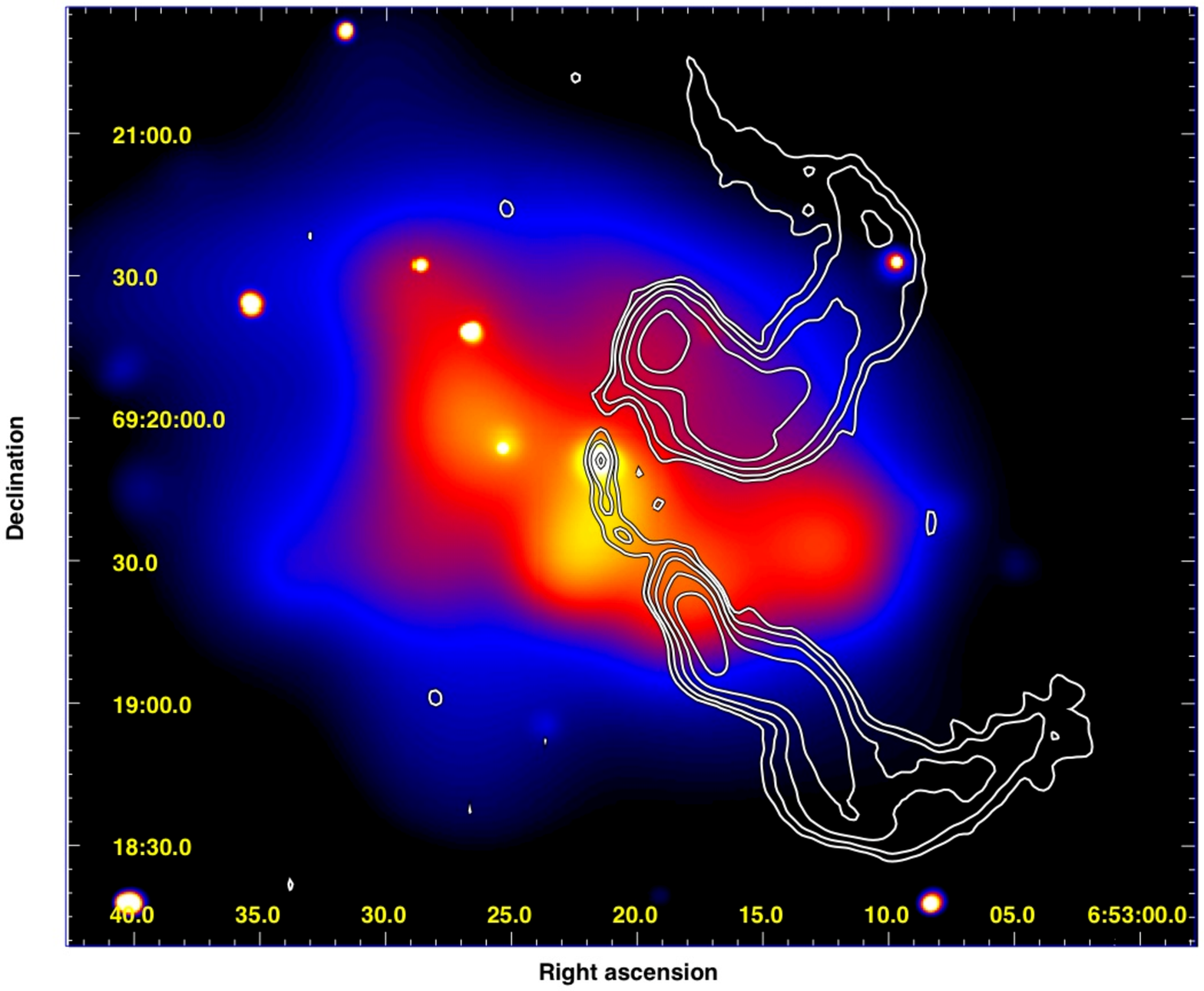}
\includegraphics{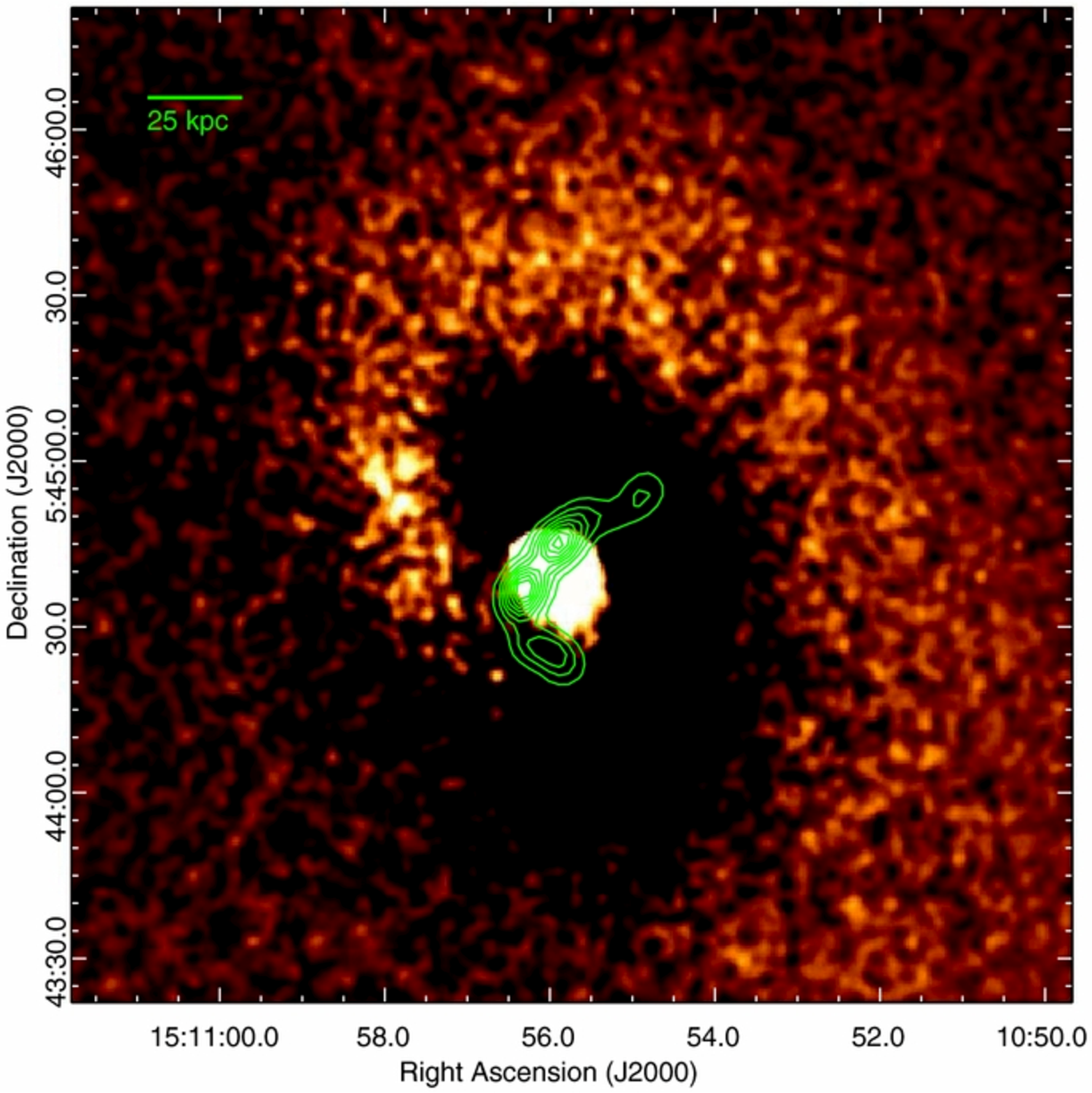}
\hspace{1.0cm}
\vspace{-1.0cm}
\end{center}
\noindent{\footnotesize
{{\it{Fig.\ 1:}} {\it Left:} Adaptively-smoothed {\it Chandra} X-ray image of A562 with radio contours superposed. A large scale
cluster-cluster merger has set the ICM in motion, providing ram pressure to bend the radio lobes
(\cite[Douglass et al. (2011)]{Douglass_etal11}).
{\it Right:} Residual {\it Chandra} X-ray image after subtracting a $\beta$-model of the canonical cool core cluster A2029 with
radio contours superposed 
(\cite[Paterno-Mahler et al. (2013)]{PaternoMahler_etal13}).
This cluster is relaxed on large scales but contains a sloshing spiral. This
gas has been set in motion from an off-axis minor merger early in the history of the cluster. This sloshing gas can bend the
radio lobes.
}}
\end{figure}

\begin{figure}[t]
\vskip 2.9truein
\begin{center}
%\hbox{
\includegraphics{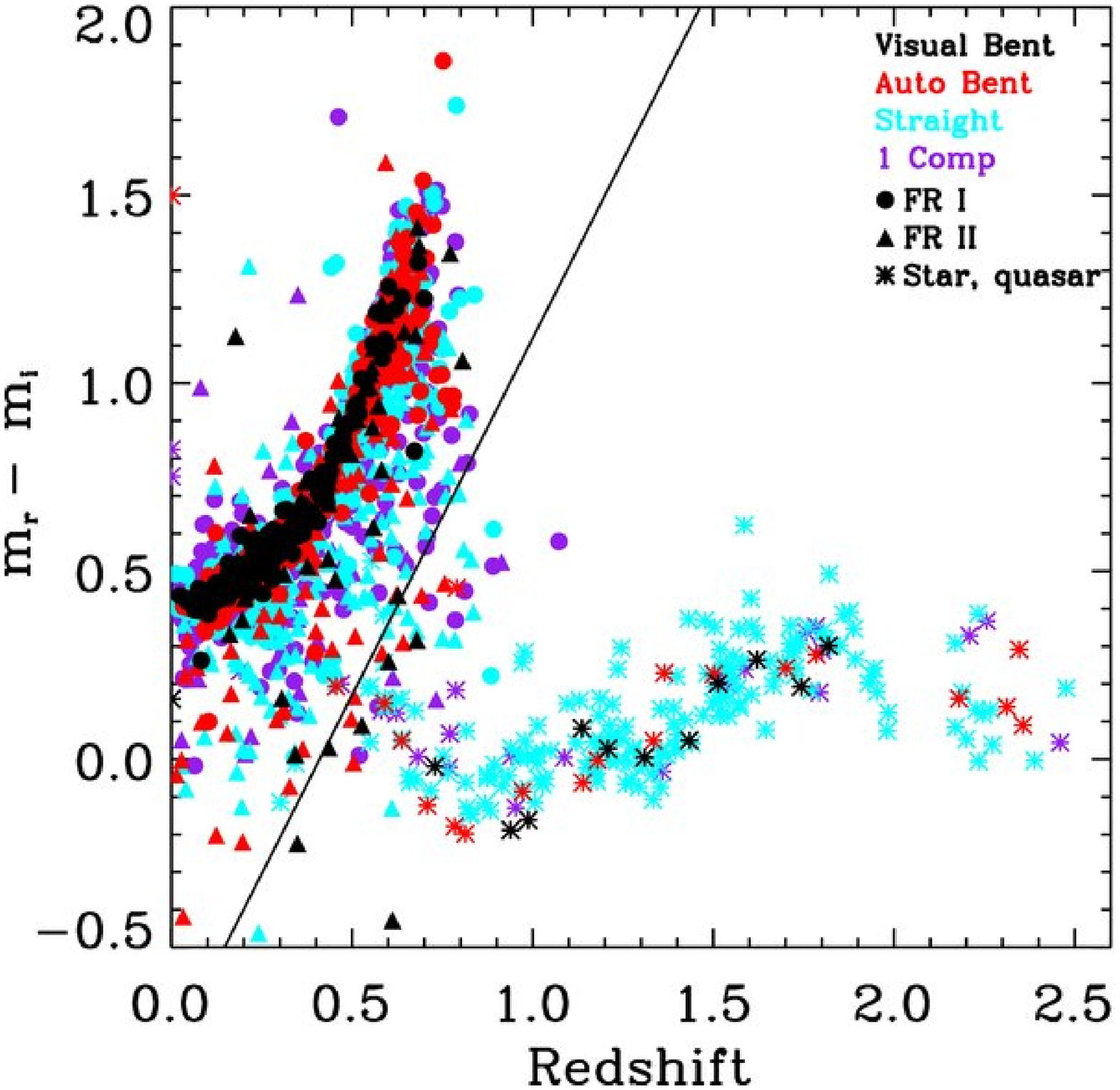}
\includegraphics{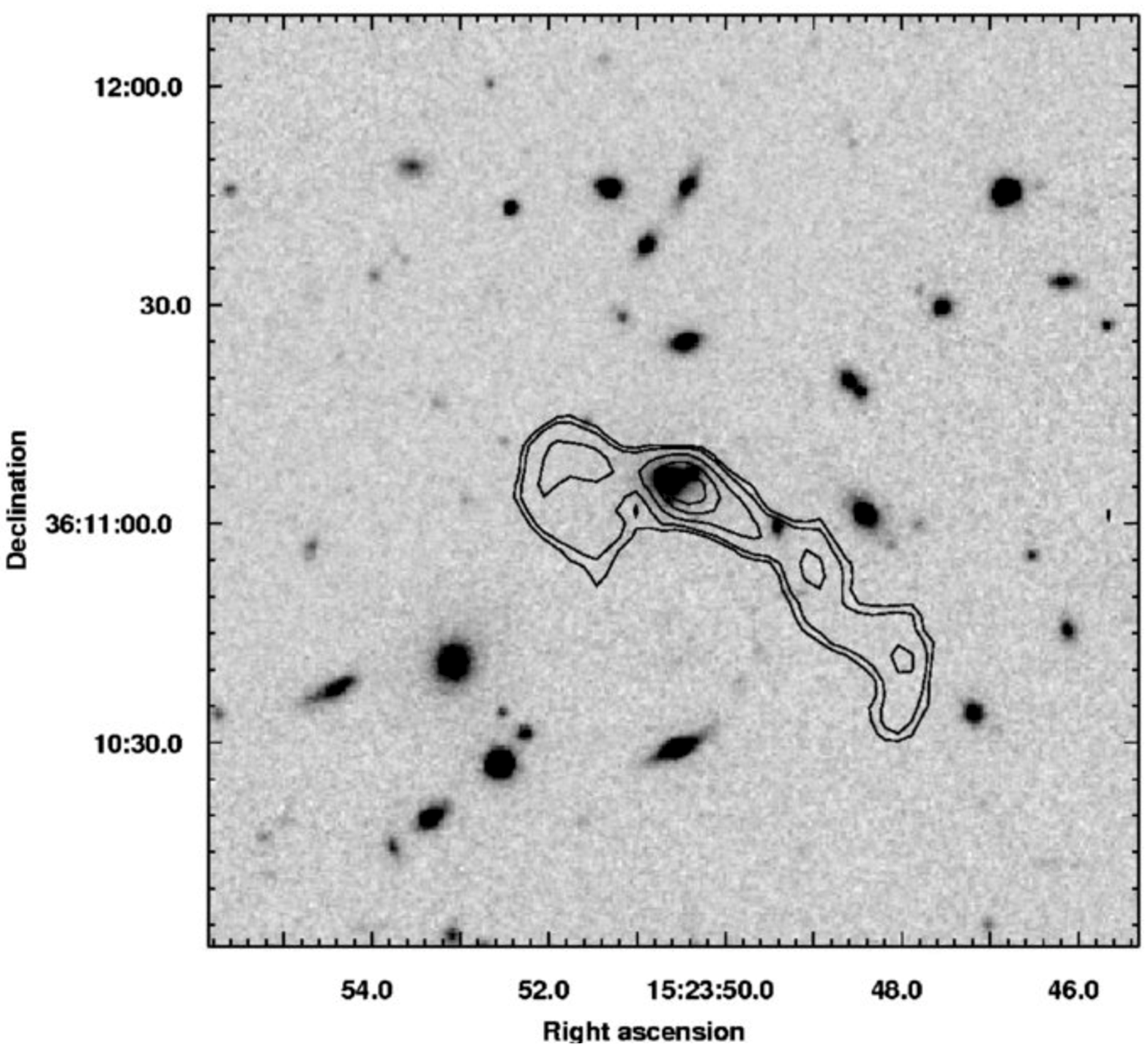}
\hspace{1.0cm}
\vspace{-1.0cm}
\end{center}
\noindent{\footnotesize
{{\it{Fig.\ 2:}} {\it Left:} Redshift vs.\ SDSS $r-i$ color for the hosts identified for the radio source samples in 
\cite[Wing \& Blanton (2011)]{Wing_11}.
The majority of the hosts have colors consistent with elliptical galaxies and are detected up to $z\approx0.7$. A small number of host objects
are quasars with emission lines allowing measurement of their redshifts to $z\approx3$ (beyond the limits displayed here).
{\it Right:} An example cluster associated with a bent-double radio source. Twenty-cm radio contours are superposed on an SDSS r-band image
of this cluster at $z=0.25$. The f.o.v.\ is 250 $\times$ 250 kpc. Adapted from \cite[Wing \& Blanton (2011)]{Wing_11}.
}}
\end{figure}

\section{The High-z COBRA Survey}

Through our work correlating radio sources from the FIRST survey with optical data from the SDSS, we identified hundreds of clusters
associated with bent-double radio sources at low-to-moderate redshift. In addition, we found 653 bent radio sources that had no optical identification in the SDSS.
Since luminous elliptical galaxy hosts would be detected in the SDSS up to $z\approx0.7$, these radio sources likely reside in clusters
at redshifts beyond this limit.
We have obtained infrared data from {\it Spitzer} as part of our Cycle 8 Snapshot program for 651 of these sources. These targets
comprise the high-z Clusters Occupied by Bent Radio AGN (COBRA) survey. All objects were observed at 3.6 $\mu$m and a subset of 139 were
additionally observed at 4.5 $\mu$m.
Based on richness measurements using our existing {\it Spitzer} data 
(\cite[Paterno-Mahler et al., in prep)]{PaternoMahler_etalprep})
we have found that approximately 200 of the sources
are associated with high-z groups and clusters up to $z\approx3$.
A candidate cluster discovered using these data is shown in Fig.\ 3. The left panel shows an overlay of the 20 cm radio contours onto
the SDSS r-band image. No host galaxy is identified, nor are surrounding galaxies. A 3.6 $\mu$m image from {\it Spitzer} is shown in the 
central panel and reveals the radio host galaxy as well as a number of surrounding galaxies. In the right panel, a deep i-band image from the 
4.3 m Discovery Channel Telescope (DCT) with the Large Monolithic Imager (LMI) is shown. Cluster galaxies are seen in this deep optical image.
Combining the infrared and deep optical data, we estimate a redshift for this cluster of $z\sim1.1$.

{\it Spitzer/IRAC} is by far the most efficient means of identifying possible high-redshift clusters associated
with the radio sources, requiring only minutes per target at $3.6~\mu m$.  Up to $z\approx3$, mid-IR observations can be used to measure the
stellar content of galaxies, and at $z>0.5$, the mid-IR probes the rest-frame NIR emission from galaxies which is the peak of the
spectral energy distribution 
(e.g.\ \cite[Krick et al. (2008)]{Krick_etal08}).
We are using deep optical observations, combined with the {\it Spitzer} observations,
to calculate photometric redshifts to high precision.  Detections of galaxy overdensities at 3.6 $\mu m$ will help us most efficiently
choose candidates for further observations in the optical, NIR, and X-ray.  We will measure the cluster stellar mass function to
$M^{*}+1.5$, and probe the distribution of galaxies on the red sequence, helping to constrain the epoch of galaxy formation and lending insight
into the history of star formation.  
Follow-up observations in the X-ray of distant, relaxed, massive clusters can be used to place constraints
on dark energy 
(\cite[Allen et al. (2004)]{Allen_etal04}).

This work was supported by the NASA Astrophysics Data Analysis Program, grant No. NNX10AC98G, NASA award RSA No.\ 1440385 issued by
JPL/Caltech, and the National Science Foundation, grant AST-1309032.
These results made use of Lowell Observatory's Discovery Channel Telescope. Lowell operates the DCT in partnership with Boston University,
Northern Arizona University, the University of Maryland, and the University of Toledo. Partial support of the DCT was provided by
Discovery Communications. LMI was built by Lowell Observatory using funds from the National Science Foundation (AST-1005313).

\begin{figure}[t]
\vskip 2.8truein
\begin{center}
%\hbox{
\includegraphics{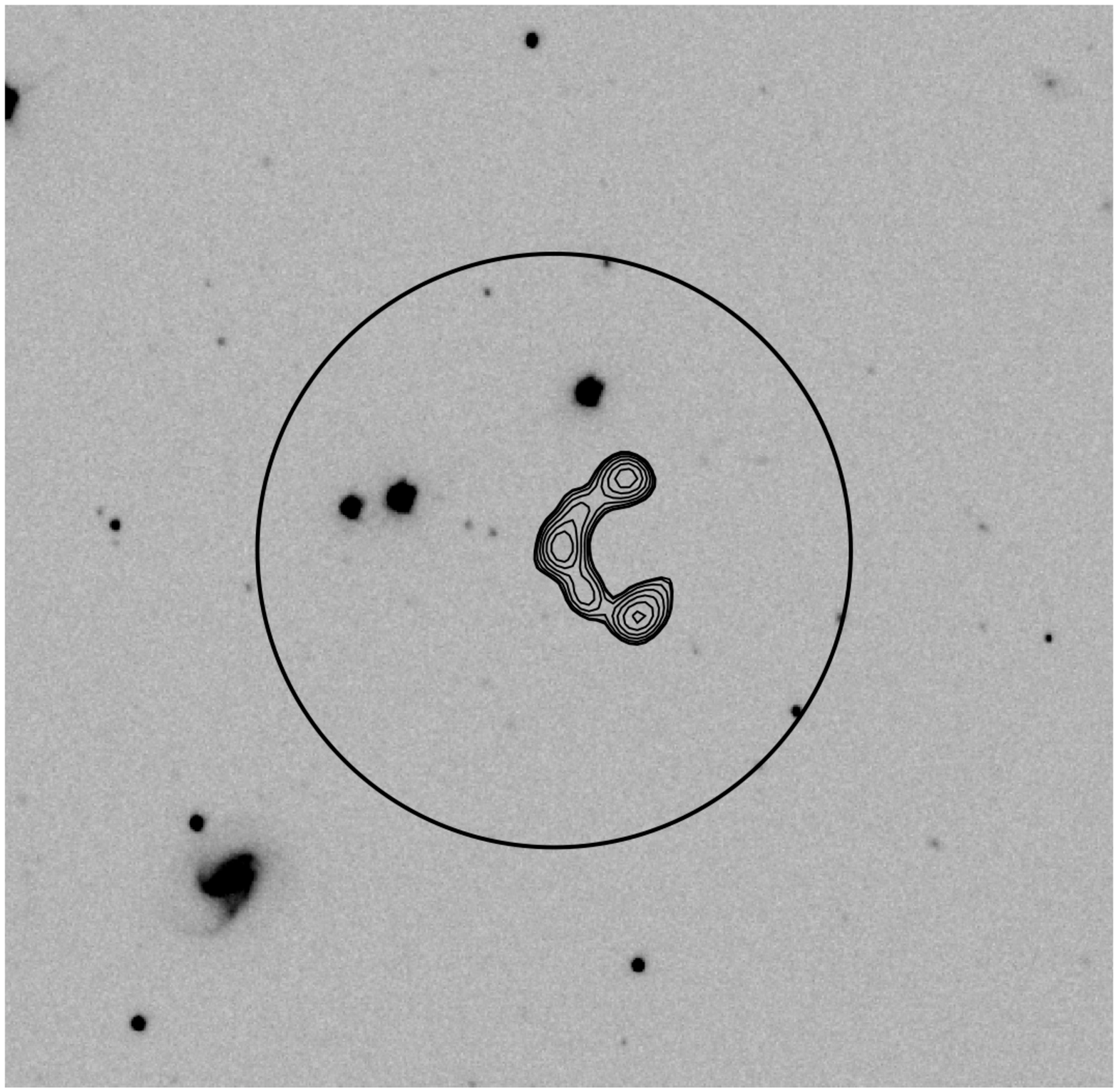}
\includegraphics{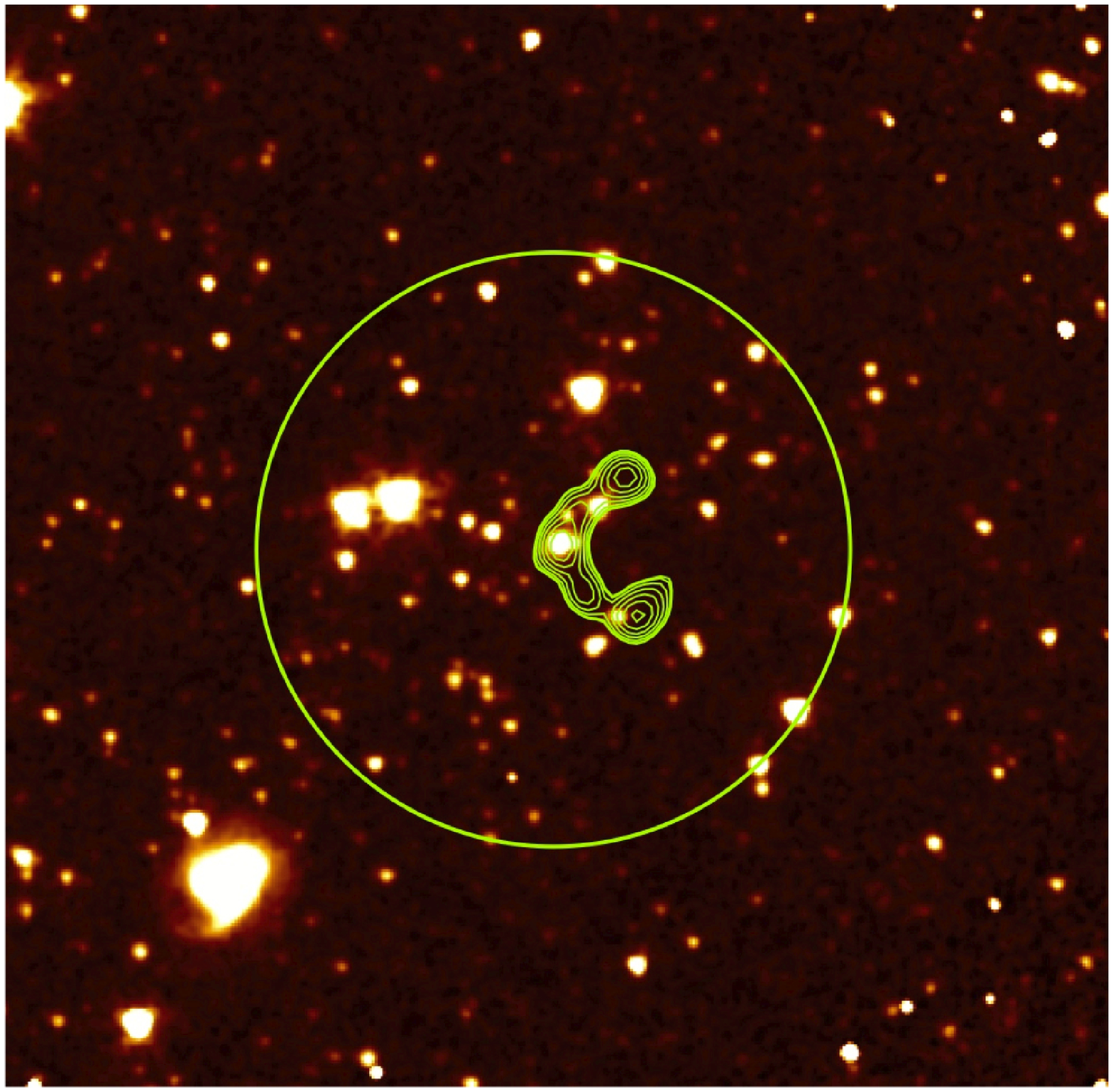}
\includegraphics{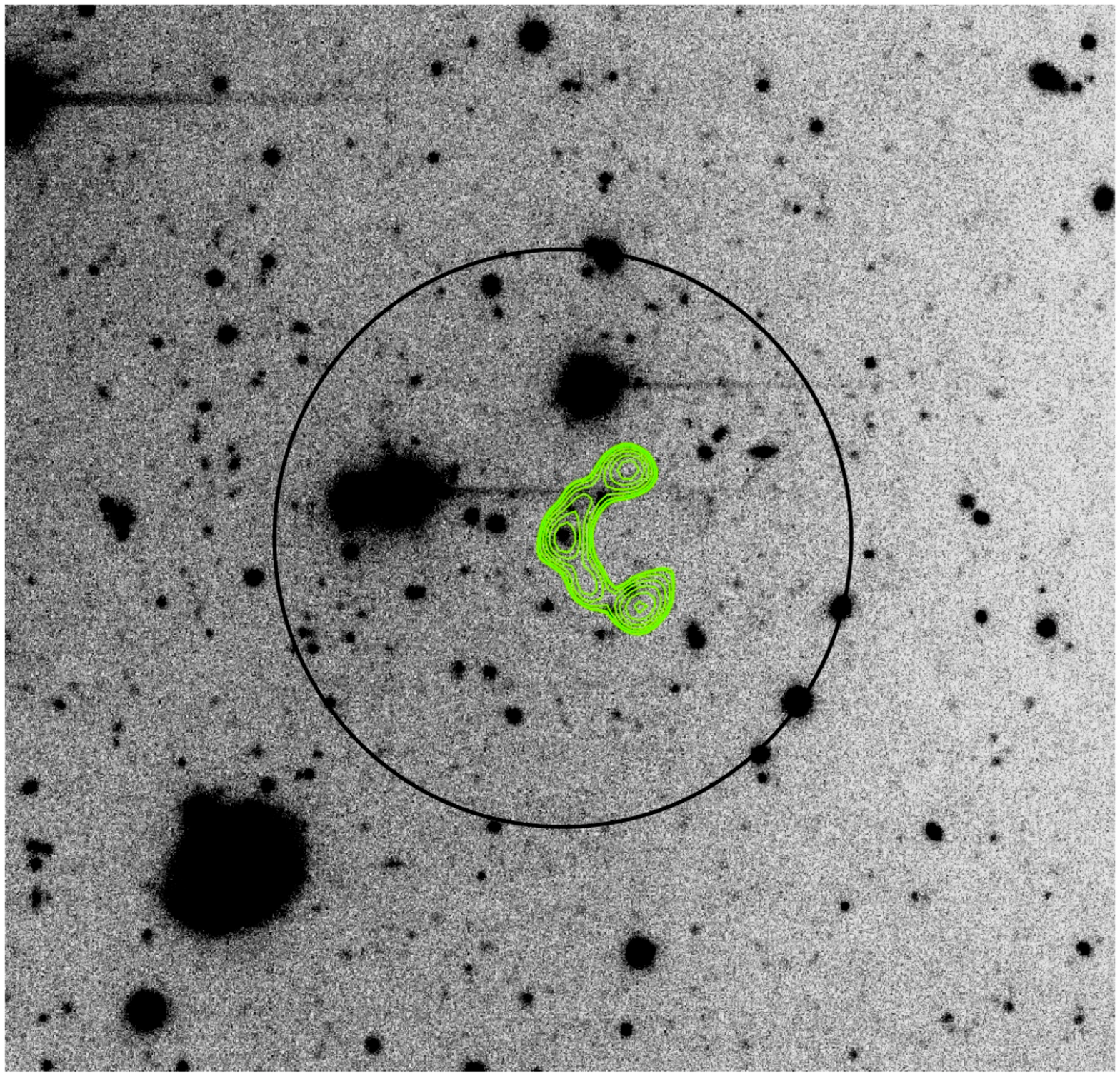}
\hspace{1.0cm}
% uncomment the following line to include your fig1b.ps postscript file
%\psfig{figure=fig1b.ps,height=7.0cm,width=7.0cm,angle=-90}
%}
\vspace{-1.0cm}
\end{center}
\noindent{\footnotesize
{{\it{Fig.\ 3:}} {\it Left:} SDSS r-band image of one of our high-z COBRA cluster candidate fields with 20 cm radio contours superposed. The radio
host and surrounding galaxies are not detected on this image. 
{\it Center:} {\it Spitzer} 3.6 $\mu$m image, showing the same 
field surrounding the bent-double radio galaxy shown in the left panel.
The infrared observations reveal the radio host galaxy as well as a large number of other galaxies, likely part of a cluster of galaxies at
high redshift. {\it Right:} DCT/LMI i-band image revealing the radio host galaxy as well as surrounding cluster galaxies that are
also detected in the IR image. Using the combined IR and optical data we estimate a redshift for this newly-discovered cluster of
$z\sim1.1$.
The circle shown has a radius of $65^{\prime\prime}$ (approximately 500 kpc at $z=1.1$).
}}
\end{figure}

\end{document}